# Exchange Biasing of the Ferromagnetic Semiconductor (Ga,Mn)As by MnO

K. F. Eid,[a] M. B. Stone,[b] O. Maksimov,[a] T. C. Shih,[c] K. C. Ku,[a] W. Fadgen,[a] C. J. Palmstrøm,[c] P. Schiffer,[a]  N. Samarth[a]1

*Department of Physics and Materials Research Institute, Pennsylvania State University, University, Park, PA, USA*

*Condensed Matter Sciences Division, Oak Ridge National Laboratory, Oak Ridge, TN, USA*

*Department of Chemical Engineering and Materials Science, University of Minnesota, Minneapolis, MN, USA*

We provide an overview of progress on the exchange biasing of a ferromagnetic semiconductor ($Ga_{1-x}Mn_xAs$) by proximity to an antiferromagnetic oxide layer (MnO). We present a detailed characterization study of the antiferromagnetic layer using Rutherford backscattering spectrometry, x-ray photoelectron spectroscopy, transmission electron microscopy, and x-ray reflection. In addition, we describe the variation of the exchange and coercive fields with temperature and cooling field for multiple samples.

---

[1] Email address: nsamarth@psu.edu

The compatibility of ferromagnetic semiconductors (FMSC) with existing semiconductor electronics [1,2] and photonics technologies [3,4] provides a strong motivation for their incorporation into potential spintronic devices. In this context, it is important to be able to exchange bias such materials to a proximal antiferromagnet (AF). The canonical FMSC $Ga_{1-x}Mn_xAs$ has been the focus of extensive experimental and theoretical studies [1,5], and is hence a natural choice for investigating both the materials science and basic physics of the exchange bias process in FMSC/AF heterostructures. We recently demonstrated the exchange biasing of $Ga_{1-x}Mn_xAs$ by MnO [6,7]. Here, we provide a more detailed overview of these experiments, including results from Rutherford backscattering spectrometry (RBS), x-ray photoelectron spectroscopy (XPS), transmission electron microscopy (TEM), and x-ray reflection (XRR). We also describe the variation of the exchange and coercive fields with temperature and cooling field measured via superconducting quantum interference device (SQUID) magnetometry for multiple samples.

Exchange bias in a ferromagnetic/antiferromagnetic bilayer system is manifested by two prominent signatures: (a) a shift in the magnetization hysteresis loop, making it centered around a non-zero magnetic field called the exchange field ($H_E$) and (b) an enhancement of the coercivity ($H_C$) of the ferromagnetic layer [8]. Since the discovery of the exchange bias phenomenon about a half a century ago [9], it has been utilized successfully in device applications [10,11]. The most important of these applications is the spin valve used in computer storage and in an array of magnetic sensor devices based on the giant magnetoresistance (GMR) effect [12]. Nonetheless, exchange bias is still not fully understood and many facets of this phenomenon remain elusive to the scientific community.

We chose MnO as the antiferromagnetic overlayer. The Neel temperature of MnO ($T_N \sim 118$ K) [13] is well within the range of attainable Curie temperature ($T_C$) of $Ga_{1-x}Mn_xAs$ ($T_C < 160$ K). Therefore, further studies may be performed to examine the effects of varying the ratio of $T_N:T_C$ or the ratio of the blocking temperature to the Curie temperature ($T_B:T_C$) through manipulating the carrier mediated ferromagnetism in $Ga_{1-x}Mn_xAs$. A special case of interest for exchange bias studies is the rarely examined limit of $T_B > T_C$ [14]. This is a unique property of the $Ga_{1-x}Mn_xAs$/MnO system and allows for more insight into the physics of exchange bias compared to more conventional exchange bias systems where typically $T_C \gg T_B$. Here, we demonstrate the exchange biasing of the $Ga_{1-x}Mn_xAs$ layer by an overgrown antiferromagnetic MnO layer both with $T_C \sim T_B$ and $T_C > T_B$.

Low temperature MBE growth is performed in an Applied EPI 930 system equipped with Ga, Mn, and As effusion cells. "Epiready" semi-insulating GaAs (100) substrates are deoxidized using the standard protocol, by heating to $\sim 580\ ^oC$ with an As flux impinging on the surface. A 100 nm thick GaAs buffer layer is grown after the deoxidization. Then, samples are cooled to $\sim 250\ ^oC$ for the growth of a 5 nm thick low temperature GaAs layer, followed by a 10 nm thick $Ga_{1-x}Mn_xAs$ layer ($x \sim 0.06$). Growth is performed under group V rich conditions with an As:Ga beam equivalent pressure ratio of $\sim 12:1$. After the $Ga_{1-x}Mn_xAs$ growth, samples are transferred *in situ* to an adjoining ultra high vacuum (UHV) buffer chamber and the As cell is cooled to the resting temperature of 110 $^oC$ to avoid formation of MnAs clusters during the subsequent Mn growth. When the As pressure in the growth chamber decreases to an acceptable level, the wafers are reintroduced into the chamber. Then, a Mn capping layer with a thickness of ~4 nm or ~8 nm is deposited. Mn growth is performed at room temperature in order to prevent interdiffusion and chemical reaction between the Mn and $Ga_{1-x}Mn_xAs$ layers [15]. Even though the capping layer is expected to be pure Mn (99.999 % source purity), the Mn overlayer rapidly oxidizes when the samples are removed from the UHV chamber.

The growth mode and surface reconstruction are monitored *in situ* by reflection high-energy electron diffraction (RHEED) at 12 keV. The thickness of the $Ga_{1-x}Mn_xAs$ layer is calculated from RHEED oscillations, while the thickness of the Mn layer is estimated from RHEED oscillations of MnAs (whose growth rate is mainly determined by the sticking coefficient of Mn) and verified using TEM, RBS, and XRR measurements. The Mn concentration in our $Ga_{1-x}Mn_xAs$ is $x \sim 0.06$, estimated from electron probe microanalysis of earlier calibration samples grown using similar Ga and Mn fluxes. The RHEED pattern during the growth of the $Ga_{1-x}Mn_xAs$ layer has a streaky 1x2 surface reconstruction suggesting the good crystalline quality of the layer. During the Mn growth, the RHEED pattern consists of sharp, elongated streaks and its symmetry is suggestive of the stabilization of a cubic phase of Mn [7,16].

Magnetization measurements are performed using a commercial SQUID. Samples are measured with the magnetic field in plane along the [110] direction as a function of both temperature and applied magnetic field. The surface and sub-surface composition is examined by XPS and RBS. The former measurements are performed using a Kratos Analytical Axis Ultra system. The photoelectrons are excited using monochromatic Al

Kα x-rays (with a photon energy of 1486.6 eV). For depth profiling, the samples are ion milled using 4 keV $Ar^+$. RBS is performed using 1.4 MeV and 2.3 MeV and 20 μC of integrated charges of $He^+$ ions with both normal and glancing angle detector geometries, corresponding to scattering angles of 165° and 108°, respectively. Both random and <100> channeling measurements are conducted to determine the composition and depth profile of the heterostructures. TEM is used to further characterize the structure of the reacted region. Cross-sectional TEM samples are prepared by chemical mechanical polishing, dimpling and ion milling using 2.7 keV $Ar^+$. The TEM is performed using a Philips CM30 transmission electron microscope under an operating voltage of 300 kV.

In order to examine the effect of post growth annealing, two protocols were designed to mount the wafers to the sample holders. In the first protocol, indium covers the entire bottom surface of the wafer. In the second protocol, only two edges of the sample are attached with indium, leaving the middle part suspended. Samples of the first kind have to be annealed at ~ 220 °C for a few minutes in order to melt the indium and remove the sample from the block. For the second type of sample, the center portion can be directly removed by cleaving without any heating, while the indium-bonded edges require a short thermal anneal. Hence, we can systematically study the effect of the short annealing incurred during removal from the wafer holders, as well as subsequent *ex situ* annealing for the identical sample. We will show that annealing has significant effects upon the capping Mn layer due to the high reactivity of Mn with oxygen.

Figure 1 (a) shows a magnetization-versus-temperature curve for a sample with $T_C$ ~ 90 K. Data are shown for two pieces from the same wafer grown using the second mounting protocol. One piece is from the indium-free portion of the wafer and is not heated after removal from the UHV chamber. Another is from the indium-bonded portion and hence undergoes a rapid thermal anneal to ~ 220 °C during sample removal. The low background magnetization at temperatures above $T_C$ indicates that the sample is of good quality without large $Mn_2As$, GaMn, or MnAs clusters. Although we observe no difference in the $T_C$ of the indium-free and indium-mounted portions of the sample, we do note that the former has a smaller low-temperature saturated moment compared to the latter. Figures 1(b)-(c) show the magnetization (*M*) of the bilayer as a function of the applied magnetic field (*H*) after the samples were cooled to the measuring temperature (*T* = 10 K) in the presence of an external magnetic field of 1 kOe. Figure 1(b) is the magnetization of an indium-free part of the wafer. The magnetization curve is symmetric about zero applied field, indicating absence of exchange bias. Figure 1(c) shows a shifted hysteresis loop measured for an indium-mounted portion of the sample. Finally, Figure 1(d) shows a hysteresis loop of an indium-free portion of the sample that was intentionally annealed in atmosphere at 200 °C for one minute. The center of the hysteresis loop is also shifted from zero. These results demonstrate that a certain amount of annealing is necessary to create exchange bias in the $Mn/Ga_{1-x}Mn_xAs$ heterostructures.

To further understand these results, we perform depth dependent XPS studies on the indium-free portion of the wafer. This is accomplished by acquiring XPS data while simultaneously sputtering away the free surface of the sample. In such measurements, time is proportional to depth below the surface. Figure 2 (a) depicts high-resolution Mn 2p spectra for the piece annealed in atmosphere at 200 °C for 1 minute. The Mn $2p_{3/2}$ line from the annealed piece is centered at ~ 641.0 eV and its position is in agreement with the binding energy of $Mn^{2+}$ indicating the formation of MnO (metallic $Mn^0$ has a $2p_{3/2}$ line at 639 eV, while lines from $Mn_2O_3$ ($Mn^{3+}$) and $MnO_2$ ($Mn^{4+}$) have binding energies of ~ 641.7 eV and ~ 642.5 eV, respectively). The two satellite lines spaced by 5.5 eV from $2p_{3/2}$ and $2p_{1/2}$ lines further support this assignment. These satellite excitations are typical for MnO and are not present in either $Mn_2O_3$ or $MnO_2$ [17,18,19]. Both their shape and position remain constant with depth, while only their intensity decreases due to a decrease in Mn content. Thus, the annealed film is nearly uniformly oxidized with MnO being the dominant form of Mn throughout the layer.

Figure 2 (b) shows high-resolution Mn 2p spectra for the as-grown piece. In contrast, the Mn $2p_{3/2}$ line from the as-grown piece exhibits a low binding energy shoulder after 60 sec. and shifts to 639 eV after 90 sec. Ar sputtering. Satellite lines also disappear at this point. This clearly indicates that while the surface layers of the as-grown piece are oxidized, metallic $Mn^0$ dominates in the bottom layers. The metallic $Mn^0$ bonded Mn would be consistent with the bottom layers being either elemental Mn or Mn in a metallically bonded compound such as MnGa or $Mn_2As$. The latter scenario is consistent with earlier studies of Mn grown on GaAs. Jin *et al* [17] reported the formation of a $Mn_2As$-type Mn-Ga-As interfacial layer during Mn growth on GaAs at 400 K. In addition, Hilton *et al* [16] found that an epitaxial $Mn_{0.6}Ga_{0.2}As_{0.2}$ layer consisting of tetragonal $Mn_2As$ and MnGa formed between Mn and GaAs as a result of solid state interfacial reactions during annealing. At first glance, it may appear surprising to propose the presence of an interfacial reacted layer even in samples that have never been heated above room temperature. Our results are however consistent with recent *in-situ* XPS studies showing that Mn growth on GaAs at temperatures as low as 95ºC leads to the formation of an 11 monolayer thick $Mn_{0.6}Ga_{0.2}As_{0.2}$ interfacial reacted layer [20].

Simulations of the random RBS spectra confirm the formation of MnO with no detectable Ga or As at the surface. The overlapped interface surface peaks for Ga and As in the glancing angle detector <100> channeling RBS spectra in Fig. 3 correspond to more Ga and As (~5 x $10^{15}$ atoms/cm$^2$) than that expected for an abrupt interface (~1-2 x $10^{15}$ atoms/cm$^2$). This clearly indicates the presence of an interfacial reacted layer. The increase in the amount of Ga and As would correspond to a ~2nm thick $Mn_{0.6}Ga_{0.2}As_{0.2}$ if there were no ion channeling in the layer. Since this reacted layer is grown epitaxially [16], some channeling may be expected, and, therefore, the reacted layer may actually be thicker.

Figure 4(a) shows a dark horizontal band (~2.3 nm thick) at the interface in the cross-sectional TEM micrograph of a sample fabricated using the first mounting protocol (complete In mounting). This sample does not show exchange bias, suggesting that the interfacial reacted layer may consist of $Mn_{0.6}Ga_{0.2}As_{0.2}$. This is consistent with the RBS channeling results of the unannealed samples which were mounted using the second protocol and which do not show exchange bias. Energy dispersive spectrometry (EDS) in the TEM confirms the surface layer as $MnO_x$ with a thickness of ~9 nm. Upon post-growth annealing in air, the samples mounted with the second protocol exhibit exchange bias and the RBS channeling interfacial Ga and As peaks increase slightly (~1 x$10^{15}$ atoms/cm$^2$). A thin bright horizontal line is observed at the interface by cross-sectional TEM [Figure 4(b)] for a sample that shows exchange bias. The change in contrast is consistent with a decrease in density and the increase in the channeled Ga and As interfacial yields with the decrease in channeling as a result of oxidation of the interfacial Mn-Ga-As layer.

Further support for the proposed reaction model comes from XRR measurements performed on a 10 nm thick $Ga_{1-x}Mn_xAs$ sample capped with a Mn layer that is nominally 10-nm thick. These samples were not exchange biased; rather the increased thickness was chosen to be able to effectively probe the grown bilayer structures with XRR techniques. Figure 5 (a) depicts the XRR spectrum for the as-grown indium-free part of the wafer (solid line). It indicates the presence of two thin layers with different electron density. The figure also shows our current attempts at fitting the XRR data (dotted line) assuming an oxide/metal/semiconductor tri-layer structure. The thickness of the oxide layer increases while that of the interfacial metallic layer decreases when the sample is annealed in atmosphere at $200^0$ C. Finally, a uniform oxide film is formed, as shown in Figure 5 (b). Figure 6 schematically shows our proposed model of the sample structure before and after post growth annealing.

We now examine the temperature and magnetic field dependent properties of the exchange bias in our $Ga_{1-x}Mn_xAs$: MnO bilayer structures. Figures 7(a)-7(d) show the hysteresis loops of exchange biased and unbiased samples. Figures 7(a) and 7(b) show that the hysteresis loop is shifted to the left or the right depending on the direction of the cooling field. In Figure 7(a) [Figure 7(b)], the sample was cooled to the measuring temperature in a field of +2500 Oe [-2500 Oe]. The hysteresis loops are clearly shifted opposite to the direction of the cooling field as is common for exchange-biased systems [8]. Ideally the zero field cooled capped sample, Figure 7(c), would be exactly centered about zero but we still see a small shift but we still see a small shift because we cannot eliminate the field coming from the magnetization of the ferromagnetic layer. In Figure 7(d) there is no shift in the hysteresis loop of the uncapped sample indicating no exchange bias. Finally, it is also important to notice that the hysteresis loops of the capped sample in Figure 7 are all wider than the loop of the uncapped sample displayed in Figure 7(d). Exchange bias is known to enhance $H_C$ of the ferromagnetic layer as well as create a shift in the hysteresis loop, $H_E$.

As the temperature of the sample is changed, $H_E$ and $H_C$ will change accordingly. Figure 8(a) shows both $H_E$ and $H_C$ as a function of temperature for a sample which has been cooled down in the presence of a magnetic field of $H = +2500$ Oe. The structure of the sample is $Ga_{0.92}Mn_{0.08}As$ (10 nm) / MnO (4 nm). Low field measurements of $M(T)$ indicate that the Curie temperature is $T_C \sim 55$ K (data not shown); $H_E$ decreases monotonically with increasing temperature until it becomes zero at $T_B = 48$ K. $H_C$ decreases, goes through a plateau around $T_B$, and then decreases monotonically to zero at $T_C$. Figure 8(b) shows the same quantities for a sample with a different $T_C$. This sample has an approximate structure of $Ga_{0.94}Mn_{0.06}As$(10 nm) / MnO(8 nm) and $T_C \sim 90$ K (see Figure 1 (a)). $H_E$ approaches zero at the same temperature as the prior sample, indicating that despite the large difference in $T_C$ for the two samples, the blocking temperature is unchanged because it depends on the antiferromagnetic layer only. Likewise $H_C$ extends beyond $H_E$ approaching zero as $T$ approaches $T_C$. We note that recent studies of Co/CoO bilayers have shown a sign reversal in $H_E$ for exchange biased systems [21]; however, we believe that the data shown in Fig. 8(b) are likely skewed by a small remnant field in the magnetometer. As we discuss in the next paragraph, small remnant fields are able to cause changes in the exchange and coercive fields in these heterostructures.

Finally, we show in Figures 9(a) and (b) the dependence of $H_E$ and $H_C$ on the cooling field for the two respective samples examined in Fig 8. Both samples show that a cooling field of only a few Oe is sufficient to create exchange bias. Only a slight change in $H_E$ is observed for cooling fields a few orders of magnitude larger than the minimum field to create bias. This is because a small field is needed to saturate the magnetization of the FMSC layer at $T_B$. The magnetization of this layer turn sets the bias; increasing the magnetic field further has no significant effect on the bilayer. When the external cooling field is small enough ($H < 7$ Oe) there is almost no exchange bias shift as expected. Alternatively, $H_C$ does not approach zero for any external magnetic field and changes very slightly with field.

In summary, we have grown a set of $Ga_xMn_{1-x}As/MnO$ heterostructures that exhibit exchange bias and an enhancement of coercivity. We have studied the dependence of the coercivity and exchange field on temperature and cooling field. Both $H_C$ and $H_E$ depend dramatically on temperature but have a much weaker dependence on the cooling field. The blocking temperature does not change from sample to sample while $T_C$ varies, most likely due to differences in $Ga_{1-x}Mn_xAs$ growth conditions. Our detailed structural studies of the capping layer indicate that it oxidizes after the removal from the UHV chamber. However, the oxidation is partial resulting in a formation of MnO/ Mn-Ga-As/ $Ga_{1-x}Mn_xAs$ tri-layer structure. Since the metallic Mn-Ga-As region does not appear to create any significant exchange bias, short annealing is necessary to uniformly oxidize this interfacial layer to form MnO. These results are important for enhancing the potential for FMSC for use in spintronics devices, the basic understanding of exchange bias, and for designing new experiments to study the optical and spin transport properties in exchange biased FMSC.

**Acknowledgements**

This research has been supported by the DARPA-SPINS program under grant numbers N00014-99-1093, -99-1-1005, -00-1-0951, and -01-1-0830, by ONR N00014-99-1-0071 and by NSF DMR 01-01318. We thank J. Shallenberger for the useful discussion and assistance with XPS measurements. We thank M. S. Angelone for the help with the XRR measurements. Work at ORNL was carried out under Contract No. DE-AC05-00OR22725, U. S. Department of Energy.

**Figure Captions:**

**Figure 1:** Magnetization as a function of temperature and applied magnetic field (hysteresis loops) for sample $Ga_{0.94}Mn_{0.06}As$ (10 nm) / MnO (8 nm) grown using second indium mounting protocol field and field cooled at $H = 1000$ Oe from $T = 200$ K to $T = 10$ K. (a) Low-field magnetization vs. temperature for two pieces from different parts of the same sample (indium-free portion (no annealing) and indium-mounted portion) measured at $H = 100$ Oe. (b) Field cooled hysteresis loop for indium-free portion of sample. There is no horizontal shift in the loop and the coercivity is low. (c) Field cooled hystersis loop for indium-mounted portion of sample. The loop is shifted and has an enhanced coercivity. (d) Field cooled hystersis loop of an indium-free free portion that was annealed at $T = 200$ °C in atmosphere for 1 minute.

**Figure 2:** Mn 2p XPS spectra acquired as a function of depth for the indium-free portion of the $Mn/Ga_{1-x}Mn_xAs$ heterostructure (a) annealed in atmosphere at 200 °C for 1 minute and (b) as-grown. Data acquired while simultaneously sputtering away the free surface of the sample using 4 keV $Ar^+$; sputtering time is proportional to depth below the free surface of the sample.

**Figure 3:** RBS channeling spectra of $Mn/Ga_{1-x}Mn_xAs$ before (solid line) and after (dotted line) post growth annealing in air using 2.3 MeV $He^+$ beams with glancing angle geometry.

**Figure 4**: Cross-sectional TEM micrographs of $Mn/Ga_{1-x}Mn_xAs$ heterostructures mounted by the first mounting protocol during growth: (a) showed no exchange bias and (b) showed exchange bias. We note that these measurements are unable to show any observable contrast between the GaAs buffer layer and the thin $Ga_{1-x}Mn_xAs$ layer due to the low Mn concentration (~6%).

**Figure 5:** XRR measurements for the indium-free portion of a Mn(10 nm)/$Ga_{1-x}Mn_xAs$(10nm) heterostructure (a) as-grown and (b) annealed in atmosphere at 200 °C for 10 minutes. Solid lines correspond to the data while the dotted lines are fits to an oxide/metal/semiconductor model described in the text. These samples were not exchange biased; the increased thickness of the AFM layer was chosen to improve signal to noise in the XRR measurement. The $Ga_{1-x}Mn_xAs$ layer was grown under similar conditions as that of exchange biased samples and so has a nominally similar Mn concentration.

**Figure 6:** Schematic drawings of Mn/$Ga_{1-x}Mn_xAs$ MBE-grown heterostructures; (a) as-grown in the MBE chamber (b) after removal from the UHV system and exposed to air and (b) after post growth annealing in air.

**Figure 7:** Hysteresis loops indicating the role of the MnO cap in producing exchange bias and the behavior of the bias with direction of the cooling field. Measurements were made at $T$ = 10 K using sample $Ga_{0.92}Mn_{0.08}As$ (10 nm) / MnO (4 nm) grown using the first indium mounting protocol. (a) and (b) loops for field-cooled measurements ($H$ = +2500 Oe and $H$ = -2500 Oe respectively). (c) Zero-field cooled hysteresis loop. (d) Field cooled hysteresis loop ($H$ = 1000 Oe) for an uncapped sample.

**Figure 8:** Exchange field, $H_E$ = -($H_{C-}$ - $H_{C+}$)/2, and coercive field, $H_C$ = (-$H_{C-}$ + $H_{C+}$)/2, as a function of temperature for field cooling at $H$ = 2500 Oe from $T$ = 200 K. (a) Sample $Ga_{0.92}Mn_{0.08}As$ (10 nm) / MnO (4 nm) grown using first indium mounting

protocol. (b) Sample Ga$_{0.94}$Mn$_{0.06}$As(10 nm) / MnO(8 nm) grown using second indium mounting protocol.

**Figure 9:** $H_E$ and $H_C$ as a function of cooling field for measured at $T$ = 10 K. Horizontal axis is plotted on two different scales and split at $H$ = 1.5 kOe. (a) Sample Ga$_{0.92}$Mn$_{0.08}$As (10 nm) / MnO (4 nm) grown using the first indium mounting protocol. (b) Sample Ga$_{0.94}$Mn$_{0.06}$As (10 nm) / MnO (8 nm) grown using the second indium mounting protocol.

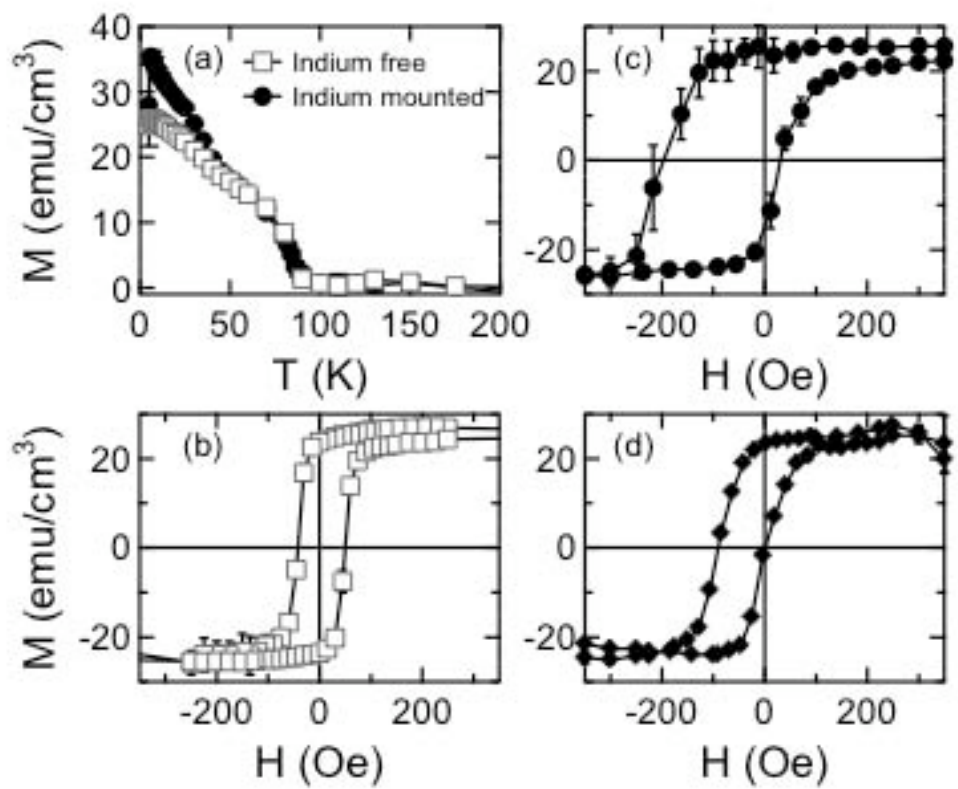

Figure 1

K. F. Eid, *et al.*

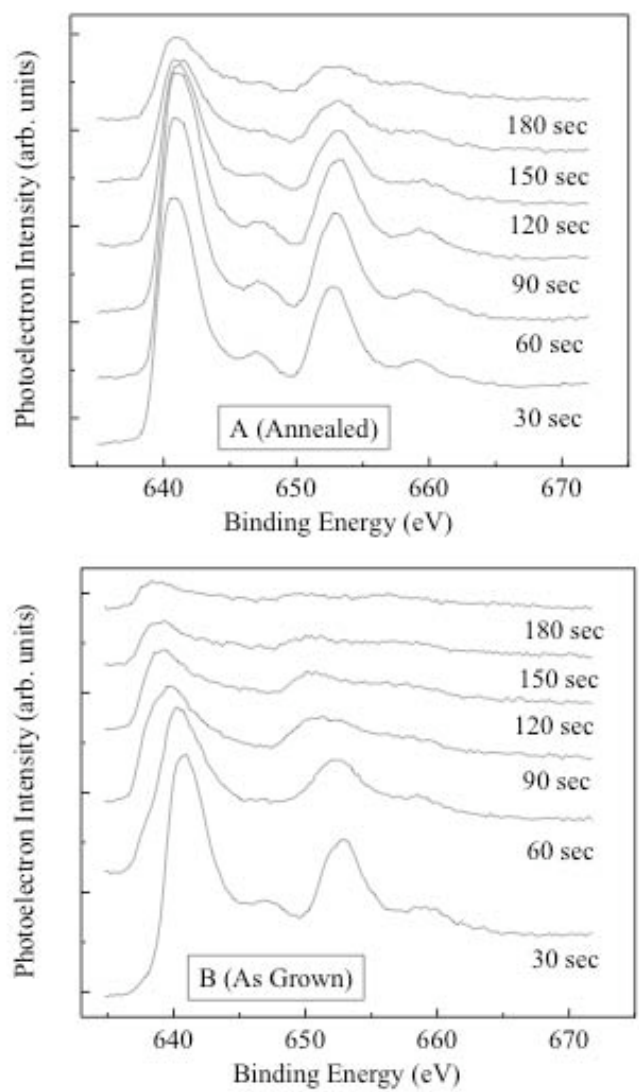

Figure 2

K. F. Eid *et al.*

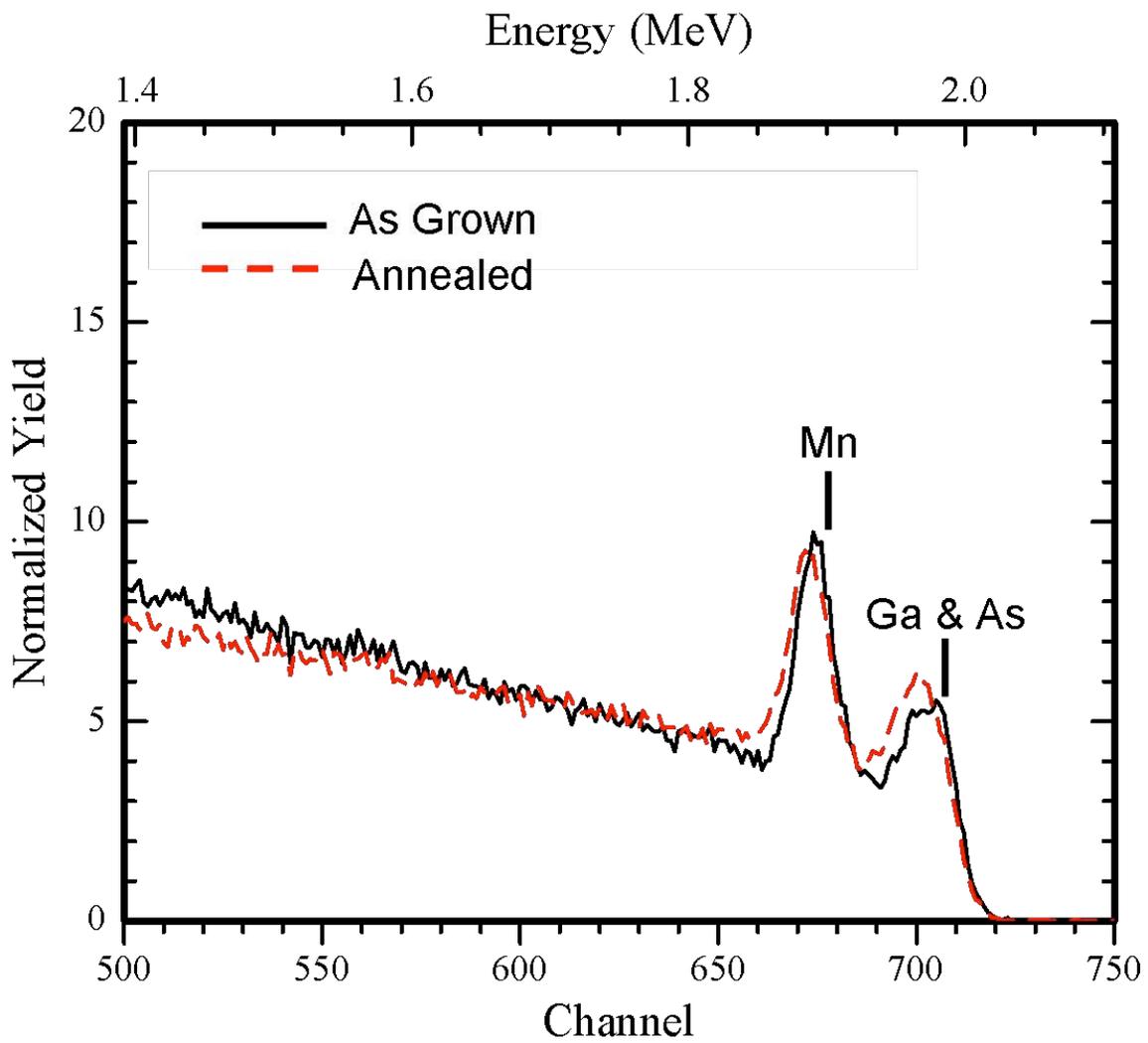

Figure 3

K. F. Eid *et al.*

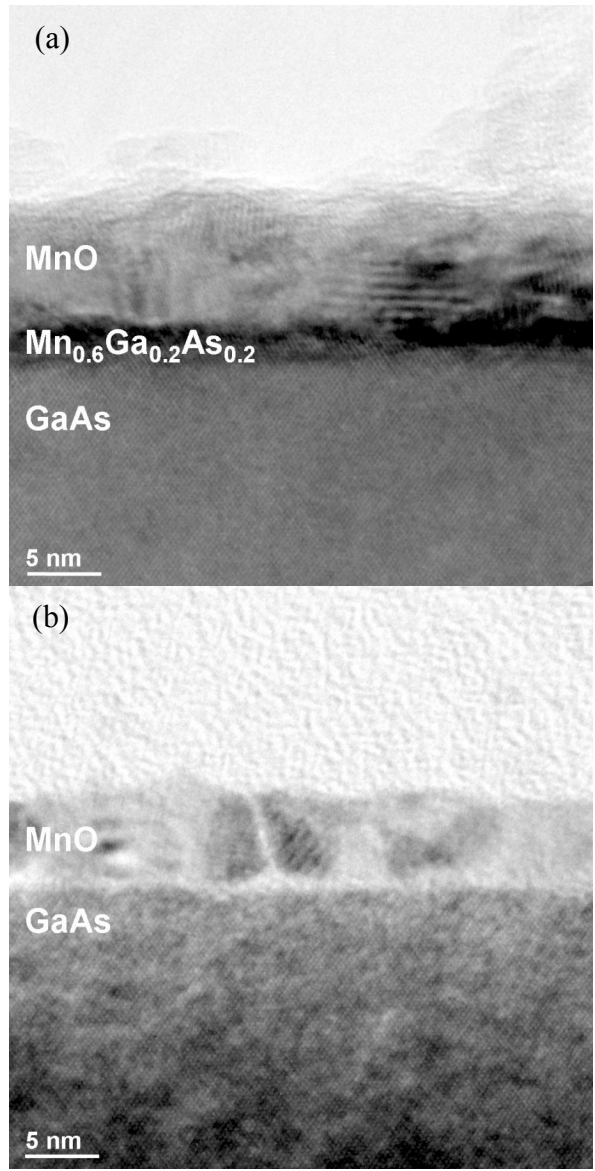

Figure 4

K. F. Eid *et al.*

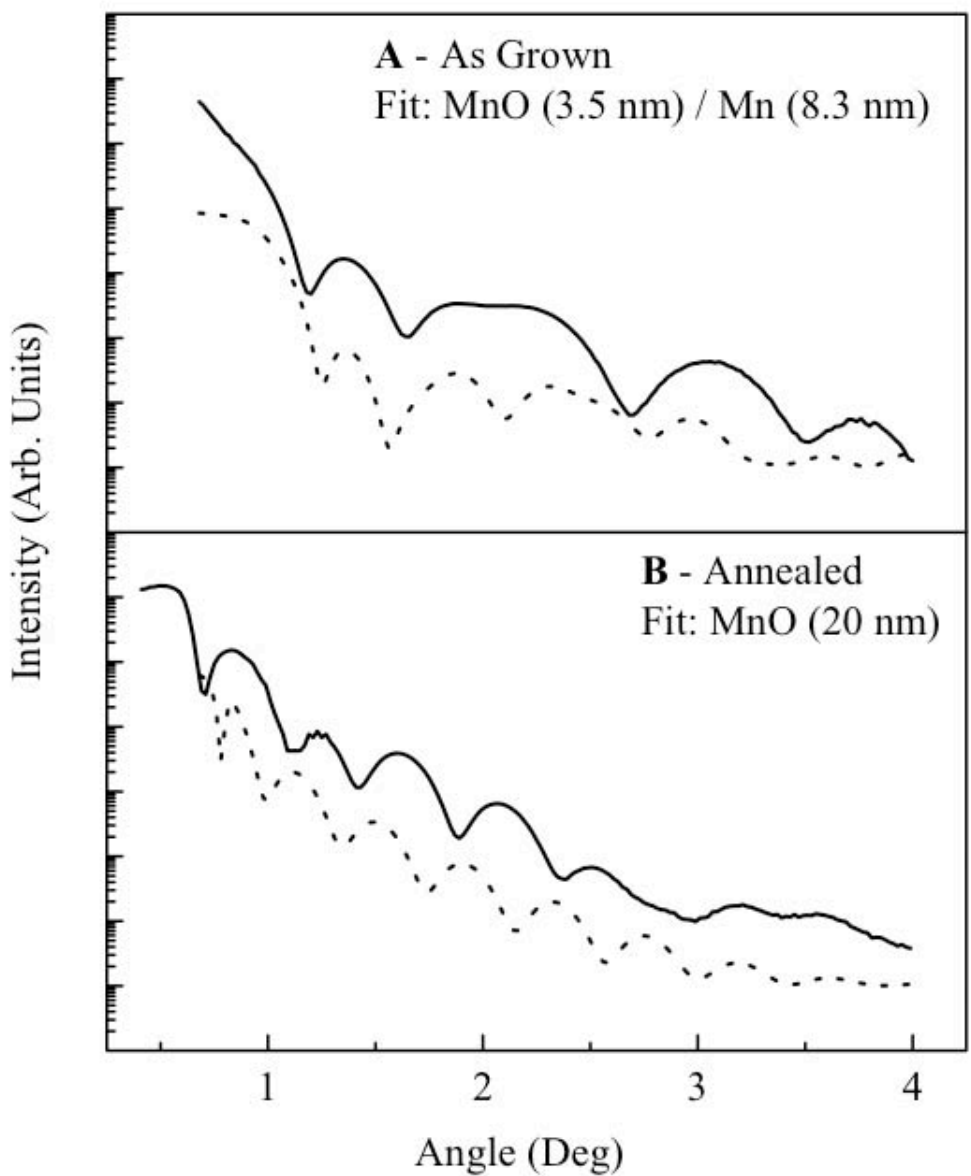

Figure 5

K. F. Eid *et al.*

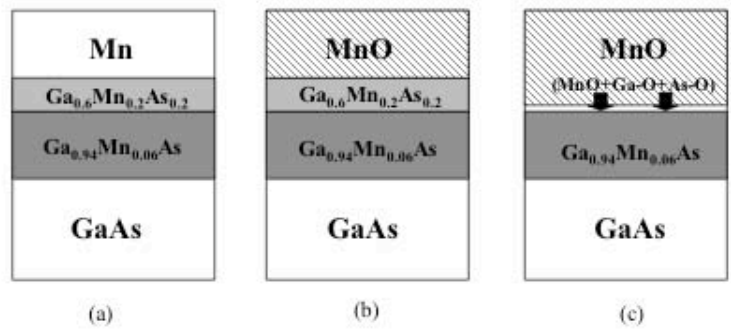

Figure 6

K. F. Eid *et al.*

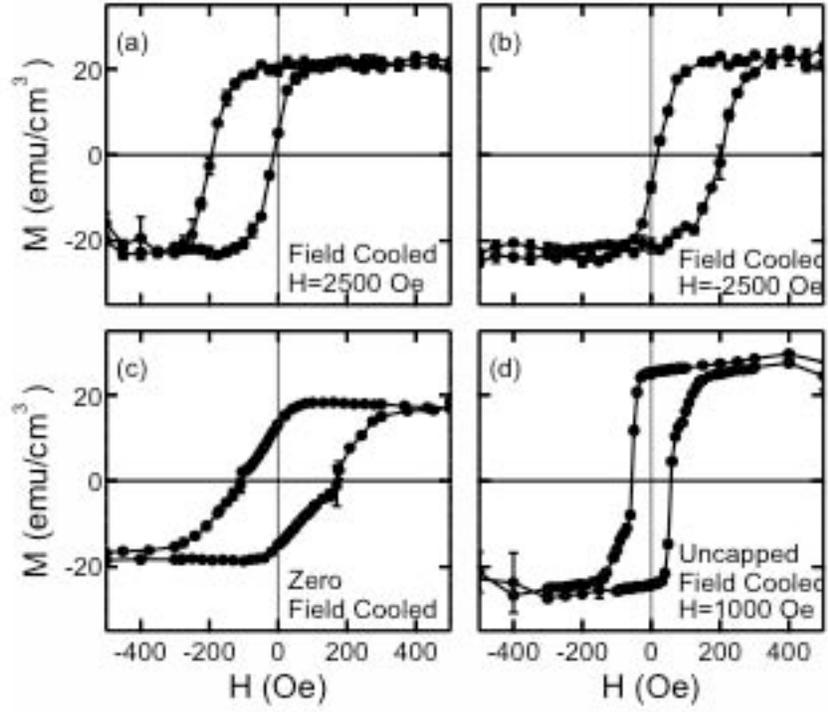

Figure 7

K. F. Eid, *et al.*

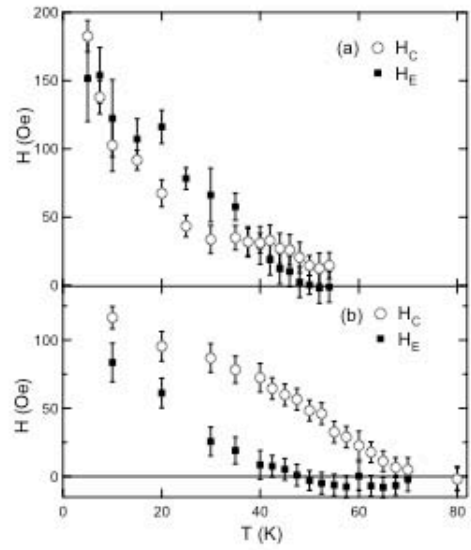

Figure 8

K. F. Eid *et al*.

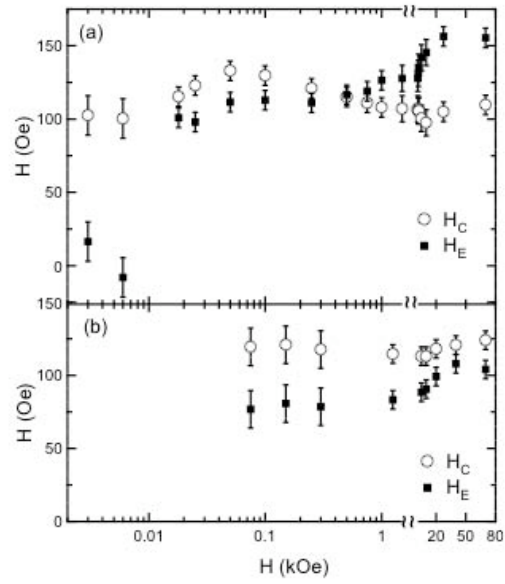

Figure 9

K. F. Eid *et al.*